\newcommand\mathmode[1]{\ifmmode {#1} \else {$#1\mkern-5mu$} \fi}

\def\ki2{$\chi^2$}

\def\b-v{$(B-V)$\kern- .15em$_{\circ}$}     
\def\md{\kern- 0.2em\raise 1.75ex \hbox{$\scriptstyle m$}
    \kern- 1em .}                           
\def\s.{\kern+ .1em\lower 0.5ex\hbox{$\buildrel ^{\prime\prime} \over
    {\rm .} \kern- .38em$}}                 
\def\d.{\kern+ .1em\lower 0.5ex\hbox{$\buildrel ^{\rm d} \over {\rm .}
    \kern- .38em$}}                         
\def\gta{\lower 0.5ex\hbox{$ \buildrel>\over\sim\ $}}
\def\lta{\lower 0.5ex\hbox{$ \buildrel<\over\sim\ $}}
\def\ss.{\kern+ .1em\lower 0.5ex\hbox{$\buildrel {\rm s}
    \over {\rm .} \kern- .05em$}}           
\def\ra#1 #2 #3 #4{$ #1^h#2^m#3\ss. #4$}    
\def\dec#1 #2 #3 {$
  #1^\circ#2\hbox{\char'023}#3\hbox{\char'175}$}    
\def\sec#1 {$ #1\hbox{\char'175}$}

\def\ls{\vskip\the\baselineskip}           

\def\msun{\mathmode{M_\odot}}

\def\lsun{\mathmode{L_\odot}}

\def\P0{\Pi_o}
\def\Teffo{T_0}
\def\Teff{T_{\rm eff}}
\def\Menv{M_{\rm env}}
\def\Mcore{M_{\rm core}}
\def\Msun{M_{\rm \odot}}

\def\d2i{{\partial^2 I_\nu\over\partial T^2}\G|_{\Teffo}}

\def\10spaces{\ \ \ \ \ \ \ \ \ \ }
\def\G#1{\Bigg#1}                           

\documentstyle[12pt,aasms4]{article}

\slugcomment{ Submitted to {\it The Astrophysical Journal (Letters)}}

\begin{document}

\title{Gravity--Mode Instabilities in Models of Post--Extreme Horizontal
Branch Stars: Another Class of Pulsating Stars?}
\author{S. Charpinet, G. Fontaine, and P. Brassard}
\affil{D\'epartement de Physique, Universit\'e de Montr\'eal, C.P.
                 6128, Succ. Centre-Ville, Montr\'eal, Qu\'ebec, Canada,
                 H3C 3J7\\ 
                 charpinet, fontaine, brassard@astro.umontreal.ca\\
                 }
\author{and\\}
\author{Ben Dorman\footnote{Also, Department of Astronomy, University of
Virginia, P.O. Box 3818, Charlottesville, VA 22903--0818}}
\affil{Laboratory for Astronomy and Solar Physics, NASA/GSFC, 
Greenbelt, MD 20771\\
Ben.Dorman@gsfc.nasa.gov\\}

\begin{abstract}

We present  new results of a stability analysis of realistic
models of post--extreme horizontal branch stars.  We find that $g$--mode
instabilities develop in some of these models as a result of a potent
$\epsilon$--mechanism associated with the presence of an active
H--burning shell.  The $\epsilon$--process drives low order and low
degree $g$--modes with typical periods in the range 40--125 s.  The
unstable models populate a broad instability strip covering the interval
76,000 K $\gta\ T_{\rm eff}\ \gta$ 44,000 K, and are identified with 
low--mass DAO white dwarfs. They descend from stars that reach the zero--age
``extended" horizontal branch with H--rich
envelope masses $\Menv\ \gta\ 0.0010$ \msun.  
Our computations indicate that some DAO stars should show luminosity variations resulting
from pulsational instabilities. We suggest looking for brightness
variations in six particularly promising candidates.

\end{abstract}


\section{ASTROPHYSICAL CONTEXT}

We have recently embarked in a systematic investigation of the
asteroseismological potential of stellar models on the extreme
horizontal branch (EHB) and beyond.  This was made possible thanks to
significant progress in our ability to compute increasingly
sophisticated and realistic models for this relatively neglected phase
of stellar evolution (see, e.g., Dorman 1995 for a review).  The models
of interest are low--mass objects ($M$ $\lta$ 0.5 $\Msun$) with outer
H--rich envelopes too thin to reach the AGB after core helium
exhaustion.  Such models cannot  sustain significant H--shell burning during
core helium burning evolution.
The core He burning phase, of typical length $\sim 10^{8}$ yr, is identified with 
subdwarf B (sdB) stars.  The stars have atmospheric parameters found in the ranges
40,000 K $\gta\ T_{\rm eff}\ \gta$ 24,000 K and 6.2 $\gta$ log $g$
$\gta$ 5.1 (see Saffer et al. 1994 and references therein).

A stability analysis of stellar models in the sdB phase of evolution has
led us to the discovery of an efficient driving mechanism due to an
opacity bump associated with iron ionization (Charpinet et al. 1996).  
It was found that both radial and nonradial ($p$, $f$, and $g$)
low order modes could be excited in some of these models.  On this basis,
we made the prediction that a subclass of sdB stars should show
luminosity variations resulting from pulsational instabilities.
The independent and exciting discoveries of the first real pulsating sdB
stars in South Africa (Kilkenny et al. 1997; Koen et al. 1997; Stobie
et al. 1997; O'Donoghue et al. 1997) gave us confidence in the basic
validity of our models.  They also led to further developments on the
observational front (Bill\`eres et al. 1997) and to refinements of the
physical description of our iron bump mechanism (Charpinet et al. 1997).

While the observational and theoretical foundations for the pulsating
sdB stars now appear to be well established, the asteroseismological
potential of their descendants, the post--EHB stars, has not been
investigated and remains of great interest.  
During the post--core--exhaustion phase  the stars contract and the
H--burning shell finally``turns on''.  With insufficient hydrogen energy
to force a red--giant star envelope, the stars live in the so--called
``AGB--Manqu\'e'' phase (Greggio \& Renzini 1990) for a period similar
to the equivalent Early (i.e., pre--Thermal Pulsing) AGB evolution phase, 
$2-4 \times 10^7 \; {\rm yr}.$ These post--EHB,
He--shell--burning models are associated with the field subdwarf O stars
(Dorman, O'Connell, \& Rood 1995 and references therein).  A majority of
these stars cluster around $T_{\rm eff}\ \sim$ 45,000 K and log $g$
$\sim$ 5.5 (Dreizler 1993).  Ultimately, the models join the white dwarf
cooling tracks near $T_{\rm eff}\ \sim$ 80,000 K and are identified, in 
the early white dwarf phase, to the low--gravity DAO white dwarfs 
(Bergeron et al. 1994).  

>From an asteroseismological
standpoint, the ignition of hydrogen at the base of the H--rich outer
layer in these hot, post--EHB models is particularly interesting.
Up to 50\% of the luminosity of the star may be provided by
hydrogen burning in a thin shell in these evolved models.  
The location of the H--burning shell strongly suggests that
a potent $\epsilon$--mechanism might drive pulsation modes there (see
below).  In contrast, the He--burning shell appears to be located too
deep at the outset to play a key role in destabilizing modes.  Moreover,
we do not expect the iron bump mechanism uncovered in sdB models to be 
relevant to post--EHB models as they are too hot (see 
Charpinet et al. 1997).

In this Letter, we report on the salient features of a stability
analysis we carried out for realistic models of post--EHB stars, and on
our discovery of pulsational instabilities in some of them.

\section{STABILITY ANALYSIS OF POST--EHB MODELS}

The equilibrium models employed in this investigation are full stellar
models taken from seven distinct evolutionary sequences.  The sequences
span the evolution from the zero--age--EHB (ZAEHB) to the cool white dwarf
phase.  The evolutionary models were computed with the methods described in 
Dorman (1992a,b) and Dorman et al. (1993), but included improvements in
the constitutive physics.  The new models use the OPAL opacities
described by Rogers \& Iglesias (1992) computed in December 1993, which 
adopted the element mix referred to as ``Grevesse \& Noels 1993.'' Where 
necessary (during He--flashes), we used new low temperature opacities by
D. R. Alexander (1995, private communication, described in Alexander \& 
Ferguson 1994) which were computed with the same element mix. These 
smoothly match the OPAL opacity set within the hydrogen ionization
zone.  The other difference in the input physics was the use of the 
Itoh et al. (1983; 1993a,b; 1994a,b) conductive opacities.

Five of the sequences (the same as those considered by Charpinet et
al. 1996) correspond to the evolution of an AGB--Manqu\'e star with a core
mass of 0.4758 \msun.  The sequences differ in that different initial 
envelope masses on the ZAEHB are considered: $\Menv =$ 0.0002, 0.0012, 
0.0022, 0.0032, and 0.0042 \msun.  Two additional sequences
 with  core masses of 0.4690 \msun and ZAEHB envelope masses
of 0.0001 and 0.0007 \msun have been added in the meantime to provide a
better mapping of the sdB region in the $\Teff$--log $g$ diagram.  In
all cases, the composition of the envelopes was assumed to be solar
($X = 0.70388$, $Z = 0.01718$), derived from calibrating a solar model
sequence to $\log L/\lsun = 0,$  $T_{\rm eff} = 5770 \; {\rm K} $ at
age 4.6 Gyr.

Figure 1 illustrates the evolutionary tracks for three of our sequences
in the $\Teff$--log $g$ plane.  For comparison purposes, the positions 
of 213 known sdB stars (according to R.A. Saffer, private communication)
are shown in the upper right region of the diagram, while the positions 
of the DAO and hot DA white dwarfs analyzed by Bergeron et al. (1994) 
are shown in the lower part of the figure.  Of prime interest in the 
present context, is the conclusion of Bergeron et al. (1994) that six of
the DAO stars in their sample (those distributed about our evolutionary 
tracks) are post--EHB objects.  In contrast, the majority of the hot 
white dwarfs they discuss have higher gravities (see  Fig. 1), 
and must be considered as post--AGB stars (see Bergeron et al. 1994). 

In the present effort, we have carried out a stability analysis for all
of our post--EHB models, i.e., those beyond the sdB phase itself.  Those
models are characterized by double He-- and H--shell burning.  We
considered all modes with $l$ = 0, 1, 2, and 3 in the 5--500 s period
window.  This was done with the finite--element nonadiabatic
pulsation code briefly described in Fontaine et al. (1994) and in Brassard, 
Fontaine, \& Bergeron (1997).  We found that, in four of our sequences, the
$\epsilon$--mechanism produced by the H--burning shell is sufficiently
potent to drive low order $g$--modes in models located at the
beginning of the white dwarf branch of the evolutionary tracks.  The
unstable models define a broad strip covering the range 76,000 K $\gta\ 
T_{\rm eff}\ \gta$ 44,000 K and are identified with the DAO phase of
post--EHB evolution.  The
thick line segments along two evolutionary tracks in Figure 1 show a
mapping of the instability strip.  The sequences with the smaller
H--rich envelope masses on the ZAEHB (i.e., those with $\Menv =$ 0.0001,
0.0002, and 0.0007 \msun) have  less active H--burning
shells throughout the evolution.  While the $\epsilon$--mechanism due to
such shells still produces significant local driving, particularly
within the instability strip, it is not strong enough to overcome
radiative damping processes in these models.

A typical unstable model is model \#32 belonging to the sequence with
$\Mcore$ = 0.4758 \msun and $\Menv$ = 0.0012 \msun, whose evolutionary
track is illustrated in Fig. 1. It has an age of $1.39 \times 
10^{8}$ yr (time elapsed since the ZAEHB), a surface gravity log $g$ 
= 7.24, a luminosity $L = 6.31$ \lsun, and an effective temperature 
$\Teff$ = 55,560 K.  Table 1 gives the periods and the e--folding times
(for the unstable modes) of the lowest order pulsation modes for this
model.  The radial modes ($l = 0$) are all stable and not listed in the
table.  In this, and in all other models showing instabilities, only the
$g$--modes are excited; the $p$--modes are never driven.  The table
indicates that the lowest order $g$--modes ($k = 1,$  2, and/or 3) are
excited and span a period range 42--123 s.  The e--folding times must be
compared with the time it takes for a post--EHB star to cross the
instability strip.  Since this time is approximately equal to $4.0
\times 10^{6}$ yr, substantially longer than most of the e--folding times, 
it would appear that the excited pulsations have plenty of time to 
develop observable amplitudes.

Figure 2 illustrates some structural properties for our representative model.
The upper panel indicates, among other things, the location of the He-- 
and H--burning shells as well as the magnitude of the nuclear energy 
generation rate there (dotted curve).  On the scale used here, the 
contribution of the He--burning shell is insignificant.  The lower panel
shows the compositional stratification as well as the luminosity profile.  The 
latter demonstrates that about 40\% of the total luminosity of the model
is due to H--shell burning at the base of the H--rich envelope.  At the 
epoch of the model, some 68\% of the hydrogen in the original ZAEHB 
envelope of 0.0012 \msun has been consumed.  Explicit tests carried out
by switching off $\epsilon_{H}$ and/or $\epsilon_{He}$ in the pulsation
calculations demonstrate, as implied by the upper panel of Figure 2, 
 that all the driving is caused by
the $\epsilon$--mechanism generated by the H--burning shell.

\section{DISCUSSION}

The $g$--mode instabilities uncovered here are very similar in nature to
the pulsational instabilities discussed by Kawaler (1988; see also
Sienkiewicz 1980) in the context of the much more luminous and hotter
central stars of planetary nebulae.  Standard models of these post--AGB
stars also show a double--shell burning structure.  As in Kawaler
(1988), the $\epsilon$--process provides here both the driving force
and a filter mechanism (selecting only a few modes as unstable), but,
unlike the case for those very luminous models, only the H--burning 
shell contributes to the process in the present situation.

The $g$--modes are favored for instability because, in the course of
post--EHB evolution, their region of formation migrates from the core in
the sdB phase (Charpinet et al. 1996, 1997) to the outer envelope in the
cool pulsating white dwarfs of the ZZ Ceti type (see, e.g., Brassard et
al. 1992).  In the post--EHB DAO phase, $g$--modes are mostly sensitive
to the physical conditions in the deep envelope, where the H--burning
shell is located.  In contrast, the He--burning shell is rather located
in the core (see Fig. 2, upper panel), is less active, and thus plays no
significant role.

The $g$--modes that are driven are those with oscillation amplitudes
that are large in the H--shell--burning region.  In practice, this means
that the largest maximum in the temperature perturbation of a mode
(usually the maximum located nearest to the surface) should nearly coincide
with the H--burning shell.  As the radial index $k$ of a mode increases,
this maximum in the temperature perturbation moves outward, ultimately
reaching a location beyond the shell.  Those modes can no longer be
efficiently driven. This then provides a filter, a cutoff in $k$, beyond
which the $g$--modes are stable.

In the upper panel of Figure 2, we have plotted the absolute value of
the Lagrangian temperature perturbation of three modes as a function of
depth in our representative model.  Note that this function is
normalized for each mode, so the only information of interest here is
related to the relative strengths and locations of maxima for a $given$
mode.  The three modes illustrated belong to the sequence with $l$ = 1.
The mode with $k$ = 2 has a period of 109.78 s and is unstable (see
Table 1).  The behavior of its temperature perturbation (solid curve)
illustrates what was written in the previous paragraph, namely that its
largest maximum nearly overlaps with the H--burning shell.  This
maximizes the efficiency of $\epsilon$--driving for that mode.  

The modes with $k \geq 4$ (and $l$ = 1) are all stable in our
representative model.  The dot--dashed curve in the upper panel of
Figure 2 shows that the largest maximum in the temperature
perturbation of the $k$ = 5 mode (a stable mode with a period of 187.84
s; see Table 1) is located well above the shell burning region, near log
q $\simeq$ --5.8, where radiative damping is important.  In this specific
case, the temperature perturbation also shows a node just below the 
H--burning shell, so the conditions are particularly unfavorable for 
efficient $\epsilon$--driving for this mode.  Finally, we illustrate 
also the behavior of the normalized temperature perturbation for a mode 
with $k$ = 8 (not listed in Table 1, but stable and with a period of 
272.70 s).  Here a secondary maximum is located right in the middle of
the H--burning shell, but the ensuing driving is overcome by damping in
the regions above where the relative amplitude of the temperature
perturbation becomes much larger.

\section{CONCLUSION}

The results presented in this paper strongly suggest that non--radial
pulsations may be present in post--EHB evolution, and 
thus the methods of asteroseismology may shed light on the
structure of stars in this phase.
We have found  that $g$--mode instabilities develop in realistic models of 
post--EHB stars.  The unstable models are the descendants of
stars with relatively massive ($\Menv\ \gta\ 0.0010$ \msun) H--rich envelopes
on the ZAEHB, and pulsate because of the presence of an active H--burning
shell at the base of the envelope.  They define a broad instability strip
covering the range 76,000 K $\gta\ T_{\rm eff}\ \gta$ 44,000 K, and are
identified with low--mass DAO white dwarfs.  Low order and low degree 
$g$--modes with typical periods around $\sim$ 80 s are predicted to be 
unstable in that instability strip.

Specifically, we suggest looking for luminosity variations in those DAO
objects already identified by Bergeron et al. (1994) as post--EHB stars.
Those are HZ 34, GD 651, Ton 353, PG 0834+501, Feige 55, and PG
0134+181.  Of course, there is no a priori guarantee that those stars
have ZAHB progenitors with envelopes in the correct mass range, since the
region of  temperature/gravity space occupied by the models with instabilities
must be quite thin.  Also,
weak mass loss (often invoked to explain the abundance anomalies of sdB
stars) may succeed in thinning that envelope to the point, perhaps, of
reducing significantly the $\epsilon$--driving.  In any case, the
proposed observations would provide interesting constraints on the amount
of hydrogen left over in these DAO stars.

\acknowledgements

This work was supported in part by the NSREC of Canada and by the fund 
FCAR (Qu\'ebec).  B.D. acknowledges support from NASA grants NAG5-700 
and NAGW-4106.

\clearpage

\begin{deluxetable}{clcclcclcc}
\tablecolumns{10}
\tablewidth{0pt}
\tablecaption{Pulsation Periods and E-folding Times for a Representative
DAO Model}
\tablehead{
\colhead{$k$} & \colhead{} & \multicolumn{2}{c}{$l=1$} & \colhead{} & 
\multicolumn{2}{c}{$l=2$}& \colhead{} & \multicolumn{2}{c}{$l=3$} \\
\cline{3-4} \cline{6-7} \cline{9-10} \\
\colhead{} & \colhead{} & \colhead{$P$ (s)} & 
\colhead{$\tau_e$ (yrs)} & 
\colhead{} & \colhead{$P$ (s)} & 
\colhead{$\tau_e$ (yrs)} & 
\colhead{} & \colhead{$P$ (s)} & 
\colhead{$\tau_e$ (yrs)}
}

\startdata
3 & & 10.99 & stable & & 10.48 & stable & &
9.78 & stable \nl
2 & & 13.63 & stable & & 11.90 & stable & &
11.24 & stable \nl
1 & & 15.61 & stable & & 15.19 & stable & &
14.75 & stable \nl
\tablevspace{1mm}
0 & & \nodata & \nodata & & 20.70 & stable & &
17.18 & stable \nl
\tablevspace{1mm}
1 & & 82.37 & $6.99\times 10^6$ & & 54.11 & $9.98\times 10^5$ & 
& 42.41 & $2.42\times 10^5$ \nl
2 & & 109.78 & $9.01\times 10^4$ & & 69.33 & $1.34\times 10^4$ & 
& 50.25 & $1.14\times 10^4$ \nl
3 & & 122.73 & $1.14\times 10^4$ & & 74.59 & $4.70\times 10^4$ & 
& 59.59 & stable \nl
4 & & 158.70 & stable & & 101.64 & stable & &
76.55 & stable \nl
5 & & 187.84 & stable & & 114.16 & stable & &
87.22 & stable \nl
\enddata
\end{deluxetable}



\clearpage
\centerline{\bf{FIGURE CAPTIONS}}

\noindent
Fig. 1 ---  Typical evolutionary tracks for EHB and post--EHB models in
the $\Teff$--log $g$ diagram.  Three tracks are illustrated, each 
corresponding to the same initial core mass, 0.4758 \msun, but with 
ZAEHB H--rich envelope masses of 0.0042, 0.0012, and 0.0002 \msun.  
The individual models along a track are represented by small crosses, 
and are joined together by dotted straight line segments.  They cover 
the evolution from the ZAEHB to the white dwarf phase.  Superimposed, in
the upper right corner, are the positions of 213 real sdB stars (small
dots).  Likewise, the positions of the known DAO (filled circles) and
hot DA (open circles) white dwarfs are illustrated in the lower part of
the diagram.  The thick segments along two of the tracks in the DAO
region correspond to the positions of unstable models driven by the 
$\epsilon$--mechanism in the H--burning shell.

\noindent
Fig. 2 ---  Lower panel: the solid curve gives the luminosity profile as
a function of fractional mass depth (log q = log $(1-M(r)/M_*)$) in our 
representative model.  The other curves show the composition
stratification.  Upper panel: the dotted curve shows the magnitude of
the nuclear energy generation rate in, and the location of, the He-- 
and H--burning shell.  The other curves refer to the amplitude of the
relative Lagrangian perturbation of the temperature for three $g$--mode
overtones with $l$ = 1.  Only the mode with $k$ = 2 is excited.

\clearpage
\begin{figure}[p]
\epsfxsize=6.0in
\leavevmode
\epsffile{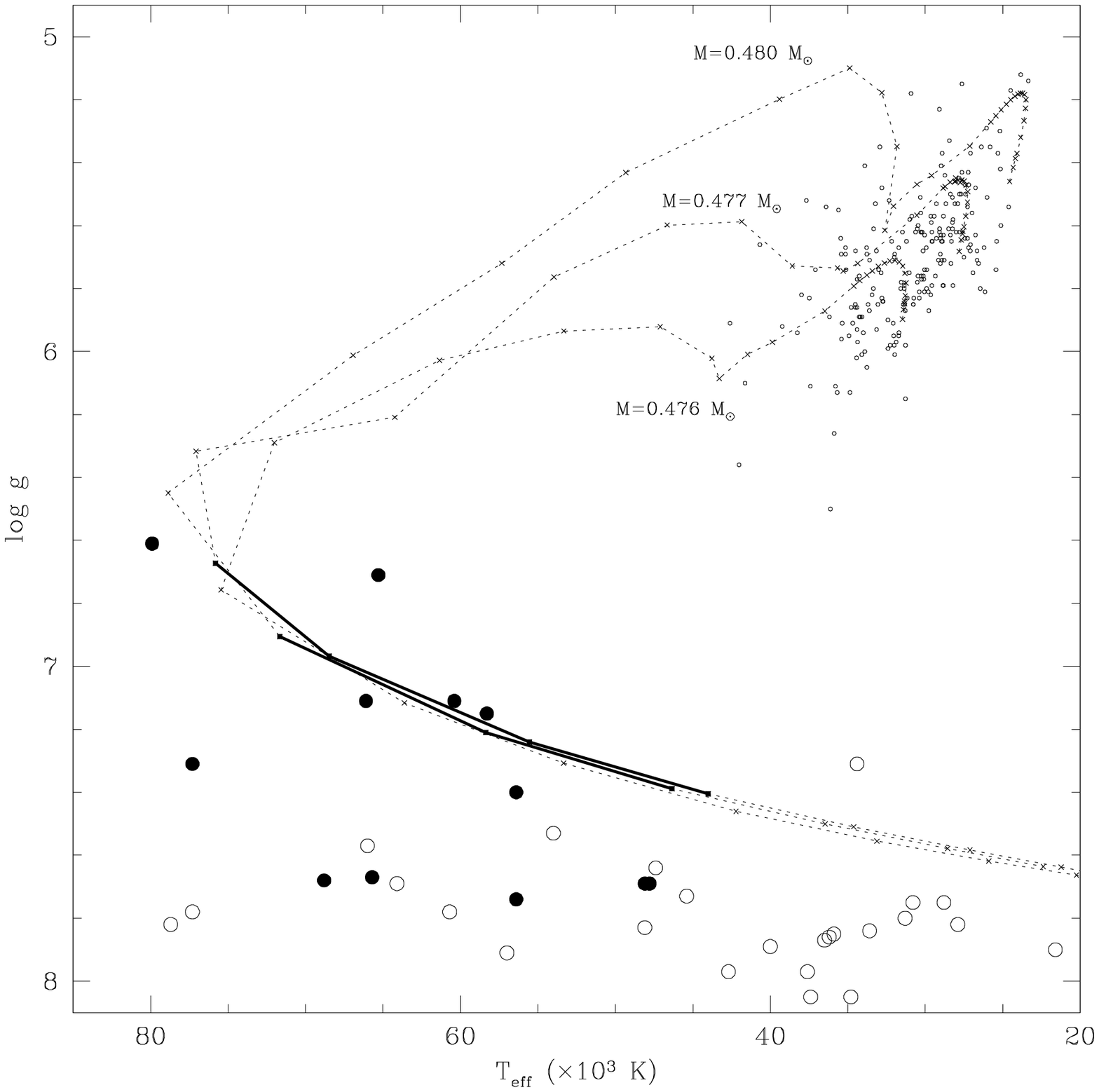}
\begin{flushright}
Figure 1
\end{flushright}
\end{figure}

\clearpage
\begin{figure}[p]
\epsfxsize=6.0in
\leavevmode
\epsffile{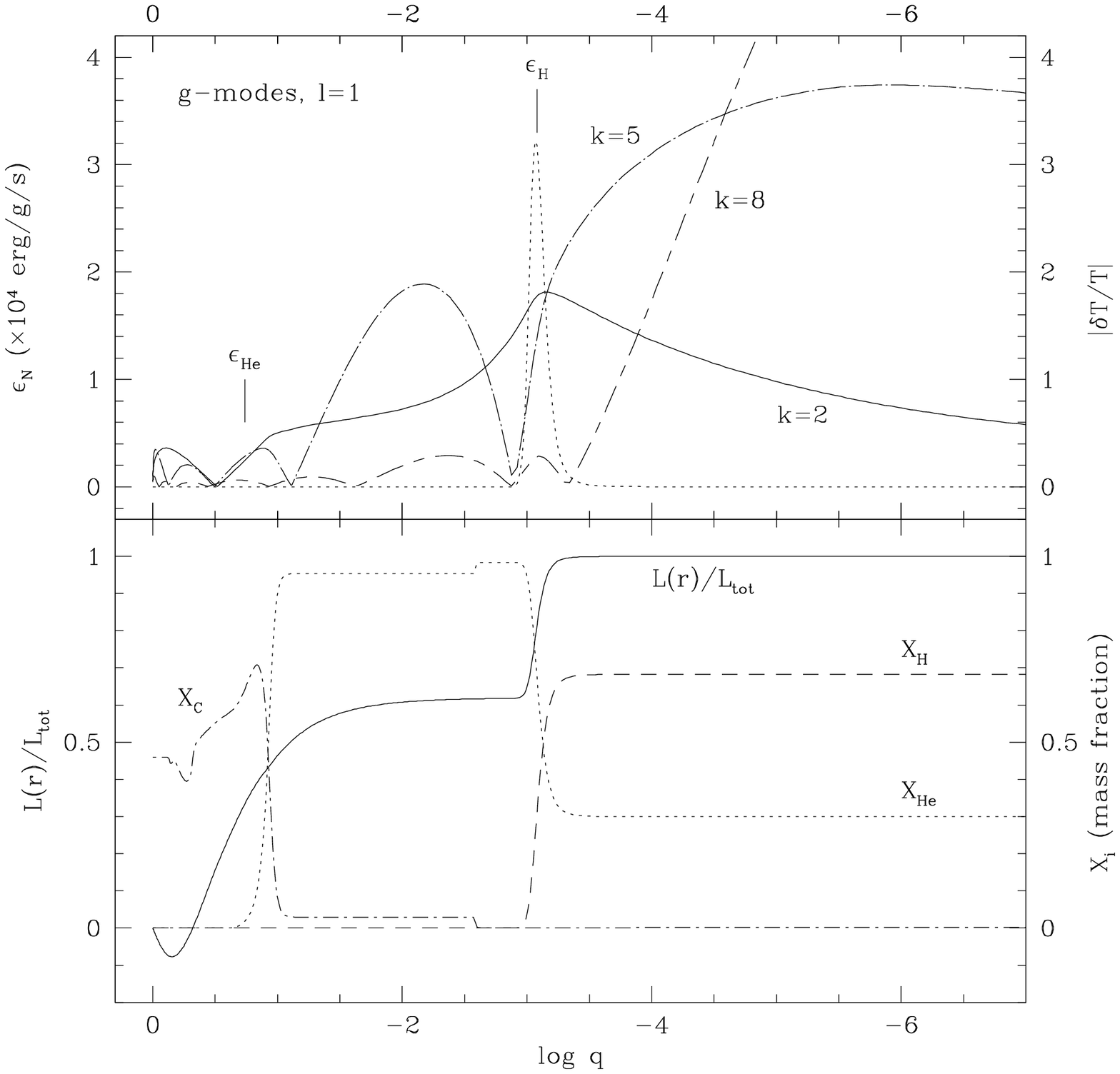}
\begin{flushright}
Figure 2
\end{flushright}
\end{figure}

\end{document}